\documentclass[10pt]{amsart}
\usepackage{geometry}                
\usepackage{graphicx}
\usepackage{amssymb}
\usepackage{epstopdf}
\usepackage{mhchem}
\usepackage[super,square,comma]{natbib}
\title{Toward the design principles of molecular machines}

\author{Aidan I. Brown}
\author{David A. Sivak}

\newcommand{\kT}{k_{\rm B}T}
\newcommand{\md}{\mathrm d}

\newcommand{\Peq}{P^{\rm eq}}
\newcommand{\Pneq}{P^{\rm neq}}
\newcommand{\Feq}{F^{\rm eq}}
\newcommand{\Fneq}{F^{\rm neq}}
\newcommand{\Plambda}{P_{\Lambda}}
\newcommand{\Plambdarev}{P_{\tilde{\Lambda}}}
\newcommand{\FO}{\text{F}_{\text{O}}}

\usepackage{titlesec}
\titleformat{\subsection}[display]{\normalfont\bfseries\centering}{\chaptertitlename\ \thechapter}{10pt}{}

\begin{document}

\maketitle
\thispagestyle{empty}

\vspace{-0.3in}\hspace{2.25in}aidanb@sfu.ca\hspace{0.5in}dsivak@sfu.ca

\begin{center}
\vspace{-0.0in}\emph{Department of Physics, Simon Fraser University, Burnaby, British Columbia, V5A1S6 Canada}
\end{center}
\vspace{0.0in}

{\bf Summary:}
Living things avoid equilibrium using molecular machines. Such microscopic soft-matter objects encounter relatively large friction and fluctuations. We discuss design principles for effective molecular machine operation in this unfamiliar context.

\section*{\uppercase{{\bf Introduction}}}

\subsection*{Life Is Fundamentally Out Of Equilibrium}
\noindent From the largest mammals down to unicellular organisms, living things manifest orderly structures, processes, and flows that are inconsistent with a state of thermal equilibrium~\cite{schrodinger44}. Cells, the micron-scale structural units of life, maintain out-of-equilibrium conditions of chemical concentrations, charge and molecular distributions, and unequal pressures~\cite{philips12,alberts14}. This ubiquity of out-of-equilibrium states flies in the face of the Second Law of thermodynamics, which pushes towards increased entropy in the absence of coordinated free energy input.

\subsection*{Life stays out of equilibrium using molecular machines}
\noindent Cells rely heavily on many types of molecular machines---macromolecular complexes that convert between different forms of energy---to achieve various tasks that create and maintain low-entropy structure.
The molecular machine adenosine triphosphate (ATP) synthase~\cite{xing05} is noteworthy, as it couples to nonequilibrium electrochemical distributions to drive the otherwise free-energetically unfavourable synthesis of ATP. Many other machines then couple to the nonequilibrium ratio of ATP to adenosine diphosphate (ADP) to drive other unfavourable reactions. \emph{E.g.}, transport motors (such as kinesin~\cite{clancy11} and myosin~\cite{sweeney10}) distribute cargoes to make them less uniform or to beat the natural timescales of diffusion. These naturally evolved machines provide inspiration for the design of synthetic molecular machines, including those that walk~\cite{bromley09}, rotate~\cite{murakami97}, and pump~\cite{cheng15} (for a discussion of natural and synthetic molecular motors, see this issue's article by Chapin Korosec and Nancy Forde).

A system at thermal equilibrium cannot have nonzero net (average) flux between different states. For example, in Feynman's ratchet~\cite{feynman66}, a wheel with asymmetric teeth engages with a pawl to prevent clockwise rotation. 
When coupled to a windmill, occasionally enough gas molecules hit the windmill to turn the wheel in the permitted counterclockwise direction. This directed mechanical work, rectifying thermal energy, appears to violate the Second Law. However, thermal fluctuations sufficient to turn the wheel are also sufficient to disengage the pawl, allowing clockwise rotation~\cite{jarzynski99}. The thermal fluctuations produce no net motion on average, hence the Second Law is preserved.

Hence molecular machines at equilibrium are not functional and do no useful work: they are as likely to hydrolyze ATP as synthesize it; as likely to transport cargoes to the left as to the right. These machines both maintain nonequilibrium conditions and themselves operate far from equilibrium, paying a free energy cost to escape thermodynamic equilibrium and achieve directed behaviour.

\section*{\uppercase{\bf Nanoscale machines run differently}}
\noindent Molecular machines and their components are nanometer-sized, and such tiny objects interact with their environment in ways that defy our physical intuition honed with macroscopic, human-sized objects.

\subsection*{Friction and inertia}
\noindent The balance of inertial and viscous (frictional) forces is quantified by the dimensionless Reynolds number $Re \equiv vL/(\mu/\rho)$, for velocity $v$, characteristic linear dimension $L$, dynamic viscosity $\mu$, and mass density $\rho$.
The micron-sized bacterium \emph{E. coli} (diameter $\sim 1 \mu$m) can swim up to $40$ $\mu$m/s~\cite{chattopadhyay06}, which in room-temperature (20$^\circ$C) water corresponds to $Re \sim 10^{-5}$. Viscous forces thus completely dominate its inertia~\cite{purcell77}, so much so that if it were to immediately stop actively swimming, it would coast only $\sim$10 pm, less than $10^{-5}$ of its own size.

Nanometer-sized molecular machines have even lower Reynolds number and less inertia. The average motion of such nanoscale objects persists only as long as something continues to `push:' they rapidly `forget' how fast and in what direction they were traveling, producing overdamped behaviour. Unlike a car motor, a molecular machine cannot rely on inertia to carry it to its next stage of operation.

\subsection*{Fluctuations}
\noindent Although low Reynolds number corresponds to overdamped motion, nanoscale objects do not require external driving forces to move. On the contrary, these nanometer-sized objects, composed of relatively soft protein material, have energy scales comparable to the thermal energy $k_{\rm B}T$ present at room temperature, so stochastic fluctuations of motion are omnipresent. Similar to the diffusive behavior of pollen grains in water~\cite{einstein05}, the components of molecular machines are constantly jostled by collisions with the surrounding medium (typically water or other proteins). Even a driven molecular machine will move in a given direction only on average, with frequent pauses and back-steps. Single-molecule experiments now directly observe this stochastic motion~\cite{isojima16}.

This overdamped motion and strong stochastic fluctuations suggest that nonequilibrium and statistical approaches are central to understanding the behaviour and design of molecular machines.

\section*{\uppercase{{\bf Nonequilibrium statistical mechanics / stochastic thermodynamics}}}

\subsection*{Driven processes}
\noindent To study a system's behaviour out of equilibrium, we can experimentally push on it. The equilibrium ensemble can be parameterized by a control parameter $\lambda$, a knob that an experimentalist can turn to change the system. Common experimental control parameters for probing molecular machines include: the distance between foci of optical traps or between an atomic force microscope (AFM) cantilever and an immobile surface, or the rotational angle of a magnetic trap. A protocol $\Lambda$ specifies a schedule for changing the control parameter from initial value $\lambda_i$ to final $\lambda_f$. For example, one can increase the distance between the foci of optical traps, thereby unfolding a biomacromolecule stretched between them (see this issue's article by Michael Woodside).

\subsection*{The Second Law at microscopic scales}
\noindent To understand the generic nonequilibrium behaviour of stochastic systems (including molecular machines), we begin with a familiar result of macroscopic physics. One form of the Second Law of thermodynamics states that the work $W$ required to drive a system via control protocol $\Lambda$ between two control parameter values and hence two corresponding equilibrium ensembles, 
\begin{equation}
W[\Lambda] \geq \Delta F_{\Lambda} \ ,
\label{eq:2ndMacro}
\end{equation}
exceeds the equilibrium free energy change $\Delta F_{\Lambda}$ between the beginning and end of the protocol $\Lambda$.
Microscopic systems, with few degrees of freedom, behave with a significant stochastic component: any particular microscopic realization may `violate' the Second Law, using less work than the free energy change. 
In this light, we realize that the Second Law is really about averages, 
\begin{equation}
\langle W\rangle_{\Lambda} \geq \Delta F_{\Lambda} \ ,
\label{eq:2ndMicro}
\end{equation}
with the angled brackets $\langle\cdot\rangle$ representing an ensemble average over many stochastic system responses to the protocol $\Lambda$. What we now recognize as the macroscopic version of the Second Law [Eq.\eqref{eq:2ndMacro}] holds for systems with Avogadro's number of degrees of freedom, where statistical fluctuations are negligible.

Research in recent decades has uncovered what are, in effect, statistical generalizations of the Second Law to microscopic systems. The Jarzynski relation~\cite{jarzynski97} places a tight constraint on nonequilibrium response through an equality between the (exponentiated) free energy change and the exponential average of the work,
\begin{equation}
\langle e^{-\beta W}\rangle_{\Lambda} = e^{-\beta\Delta F_{\Lambda}} \ ,
\end{equation}
for $\beta \equiv (k_{{\rm B}}T)^{-1}$, Boltzmann's constant $k_{{\rm B}}$, and environmental temperature $T$. The statistical Second Law [Eq.\eqref{eq:2ndMicro}] is immediately recovered upon applying Jensen's inequality~\cite{Cover:2006:Book}, which states that for a convex function $f$ and random variable $Y$, $\langle f(Y)\rangle \geq f(\langle Y\rangle)$.

\subsection*{Statistics of microscopic reversibility}
\noindent In systems satisfying microscopic reversibility~\cite{casimir45}, for any forward trajectory the time-reversed trajectory is also possible.
A further generalization of the Second Law under conditions of microscopic reversibility is the Crooks fluctuation theorem~\cite{crooks98, crooks99},
\begin{equation}
\label{eq:crooks}
\frac{\Plambda(X)}{\Plambdarev(\tilde{X})} = \exp\left[\beta\left(W[X|\Lambda] - \Delta F_{\Lambda}\right)\right] \ ,
\end{equation}
where $\Plambda(X)$ is the probability that the system follows trajectory $X$ during the forward protocol $\Lambda$, $\Plambdarev(\tilde{X})$ the probability for the time-reversed trajectory $\tilde{X}$ during the reverse protocol $\tilde{\Lambda}$, and $W[X|\Lambda]$ is the work used to drive the system along trajectory $X$ during the forward protocol. This places an even stronger constraint on nonequilibrium behaviour: \emph{for each trajectory} there is a fixed ratio between the probability of a forward trajectory in response to the forward protocol, and the probability of the time-reversed trajectory for the reverse protocol. Rearranging and averaging over trajectories, given a particular protocol, recovers the Jarzynski relation.

These so-called fluctuation theorems relate reversibility to work and free energy changes.
More general nonequilibrium contexts feature multiple distinct nonequilibrium drives beyond an external work source: multiple reservoirs of temperature and/or chemical potential, out-of-equilibrium boundary conditions such as shearing, etc.  
For such systems, the sufficient statistic governing reversibility is entropy production (in the system and environment)~\cite{seifert12}. Here the more general form of the Crooks theorem is
\begin{equation}
\frac{\Plambda(X)}{\Plambdarev(\tilde{X})} = \exp\left(\frac{\Delta S[X]}{k_{{\rm B}}}\right) \ ,
\end{equation}
with entropy production $\Delta S[X]$ during the forward trajectory $X$.

Recent research seeks to apply these fundamental theoretical advances in the context of biomolecular energy and information conversion. What limits are imposed on core biological processes by basic physical considerations?  What design principles describe systems that saturate/reach those limits?  Are these principles manifested in machines evolved by nature?  And do they give useful insights for designing novel machines?  

\section*{\uppercase{{\bf Optimal control}}}
\noindent Just as we judge the efficiency of a car engine compared to theoretical limits (the Carnot limit or finite power equivalents), how do evolved and synthetic molecular machine efficiencies compare with the physical limits for stochastic nonequilibrium energy transduction? More abstractly, how can one drive a system from one equilibrium ensemble to another, in a fixed timespan, with the least energetic effort (work)?

\subsection*{Theory}
\noindent For a control parameter changing sufficiently slowly, the required instantaneous input power
is~\cite{sivak12} 
\begin{equation}
\label{eq:power}
\mathcal{P}(t_0) = \left(\frac{\md F}{\md\lambda}\frac{\md\lambda}{\md t}\right)_{t=t_0} + \zeta(\lambda(t_0))\cdot\left(\frac{\md\lambda}{\md t}\right)^2_{t=t_0}\ .
\end{equation}
The first term represents the equilibrium free energy change, the work done in the quasistatic limit when the system remains equilibrated throughout the protocol. The second term represents the excess power, the extra energy required due to the system being out of equilibrium, and it depends on a generalized friction coefficient $\zeta$ in the space of control parameters. 

$\zeta$ governs near-equilibrium response and represents the energetic cost of changing the control parameter sufficiently fast to drive the system out of equilibrium. It is a function of equilibrium system fluctuations,
\begin{equation}
\label{eq:friction}
\zeta(\lambda(t_0)) \equiv \beta\int_0^{\infty} \md t\langle\delta f(0)\delta f(t)\rangle_{\lambda(t_0)} \ ,
\end{equation}
where $\delta f \equiv f(t_0) - \langle f\rangle_{\lambda(t_0)}$ is the deviation of the conjugate force $f$ from its equilibrium average $\langle f\rangle_{\lambda(t_0)}$ at fixed control parameter value $\lambda(t_0)$. For instance, when the control parameter is the optical trap position, the conjugate force is the tensile force with which the biomolecule resists further extension. $\langle\delta f(0)\delta f(t)\rangle_{\lambda(t_0)}$ is an autocorrelation function: at $t=0$ it equals the force variance, while for $t>0$ the autocorrelation function represents how quickly the system forgets its initial condition.
Eq.~\ref{eq:friction} is an example of a fluctuation-dissipation theorem relating equilibrium fluctuations to dissipation out of equilibrium. 

How do we minimize this extra energy to waste as little as possible? In Eq.~\ref{eq:power}, the friction coefficient [Eq.~\ref{eq:friction}] expresses the rate of excess work accumulation along a control protocol -- this friction coefficient defines a metric providing a measure of path length in control parameter space. Protocols that minimize dissipation follow geodesics (shortest paths) in the curved space defined by the friction coefficient metric. This is entirely analogous to minimization of airline flight distances along great circle routes: though on a Mercator projector these routes look curved, they are in fact shortest paths on the curved surface of the earth. 

Along such optimal paths, the optimal protocol proceeds such that the excess power is constant:
\begin{equation}
\label{eq:optimal}
\frac{\md\lambda^{\text{opt}}}{\md t} \propto \frac{1}{\sqrt{\zeta(\lambda(t))}} \ .
\end{equation}
We have used this framework to examine optimal protocols in model systems~\cite{sivak12,zulkowski12,zulkowski13,sivak16}. 

\begin{figure}[tbp] 
	\centering
	\hspace{-0.3in}
	\begin{tabular}{c}
		\hspace{0.200in}\includegraphics[width=2.375in]{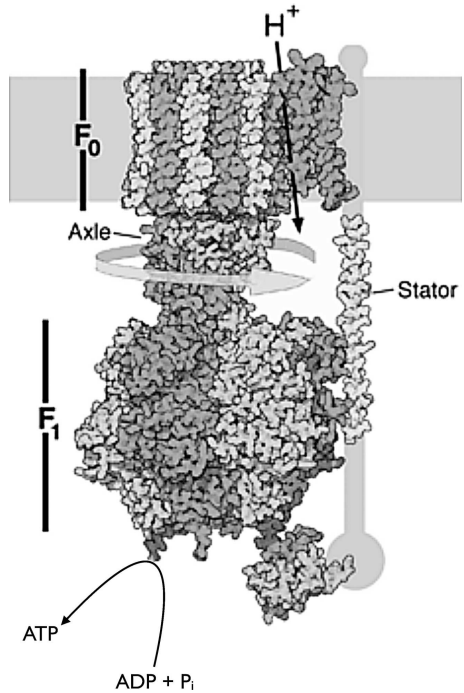}\\
	\end{tabular}
	\caption{\label{fig:atpsynthase} 
{\bf ATP synthase,} a composite rotary motor. The top motor, $\FO$, is embedded in the membrane and couples proton flow across the membrane (down its electrochemical gradient) to rotation of the central axle or crankshaft. The bottom motor, $\text{F}_1$, couples this rotation to synthesis of ATP from ADP and phosphate. Image adapted from the Protein Data Bank~\cite{pdb00,rastogi99,gibbons00,wilkens05,delrizzo02}.}
\end{figure}

\subsection*{Ramifications for ATP synthase}
\noindent ATP synthase couples transport of hydrogen ions down their gradient to synthesis of ATP from ADP and phosphate, against a chemical potential difference favouring ATP hydrolysis. Though ATP synthase is a large and intricate molecular complex, communication is mediated through a relatively simple mechanical coordinate, the rotational angle of a crankshaft connecting the integral membrane $\FO$ subunit to the soluble $\text{F}_1$ subunit (Fig.~\ref{fig:atpsynthase}). Single-molecule studies of ATP synthase typically excise the $\FO$ subunit and attach an experimental handle (e.g., a magnetic bead) to the crankshaft, and monitor or force rotation using a magnetic trap. Such experiments suggest that $\text{F}_1$ can approach near 100\% efficiency~\cite{kinosita00,toyabe11}.

Experimental observation of $\text{F}_1$ rotational statistics indicates a small number of metastable angular states separated by energetic barriers. When the magnetic trap is centered at an energetic barrier, equilibrium probability is equally split between the adjacent metastable states, giving maximal force variance $\langle \delta f^2 \rangle$ and maximal force relaxation time, hence maximizing their product, the friction coefficient. Eq.~\ref{eq:optimal} provides intuition on how an experimentalist (or $\FO$ \emph{in vivo}) should drive rotation to minimize energy expenditure: where the friction coefficient is large---where the system puts up large resistance to rapid control parameter changes, at the rotational energetic barriers---the minimum-dissipation protocol proceeds slowly, giving thermal fluctuations maximal time to kick the system over the barrier `for free.' For a double-well potential, a protocol optimized according to Eq.~\ref{eq:optimal} can dissipate less than half the energy of a naive (constant-velocity) protocol~\cite{sivak16}.

\section*{\uppercase{{\bf Nonequilibrium free energy}}}

\noindent In general, driving a system out of equilibrium costs energy, beyond the system free energy change $\Delta F^{\text{eq}}$. What happens to this extra energy put into the system, and how much of it can be used to do useful work?

The system free energy, in or out of equilibrium~\cite{sivak12b}, can be defined as $F = \langle E\rangle - TS$, with average energy $\langle E\rangle = \sum_xP(x)E(x)$ and entropy $S = -k_{\text{B}}\sum_x P(x)\ln P(x)$, for system microstates $x$.
At the conclusion of a nonequilibrium protocol $\Lambda$,
the system will generally be out of equilibrium. 
The final nonequilibrium distribution $\Pneq_{\Lambda}(x)$ over microstates will generally have a larger free energy $\Fneq_{\Lambda}$ than the equilibrium ensemble $\Peq_{\lambda}(x)$ corresponding to the final control parameter value $\lambda$. 
As the system relaxes to equilibrium from this nonequilibrium state, it can 
do work on a coupled mechanical system. 
The available work is the difference between these nonequilibrium and equilibrium free energies,
\begin{equation}
\Fneq_{\Lambda} - \Feq_{\lambda} = \kT \, D(\Pneq_{\Lambda} \| \Peq_{\lambda}) \ ,
\end{equation}
which is proportional to a central information-theoretic quantity, the relative entropy (Kullback-Leibler divergence)~\cite{Cover:2006:Book} $D(P_1 \| P_2) \equiv \sum_xP_1\ln[P_1(x)/P_2(x)]$ between the nonequilibrium and equilibrium probability distributions.
Full probability distributions require vast numbers of samples to measure experimentally, but using the Crooks theorem this relative entropy can be estimated from measurements of mean work~\cite{sivak12b}.

Molecular machines couple nonequilibrium reactions, and thus as part of their core function may utilize this nonequilibrium free energy, say by driving a downstream process using the transiently available free energy following activation by an upstream process.

\section*{\uppercase{{\bf Tradeoffs of dissipation and time symmetry breaking}}}
\noindent Eliminating energy waste cannot be the only goal -- a perfectly efficient process (dissipating no energy) would be completely reversible, and hence unable to achieve directed behavior. The Second Law says that irreversible operation requires \emph{some} energy loss, but is silent about quantitative details.

Irreversibility, which can be thought of as the `length of time's arrow'~\cite{feng08}, can be operationalized through the intuitive question: if I reverse the flow of time, how different will my dynamics look? More quantitatively, how different is the ensemble of time-reversed trajectories from the more familiar forward-time trajectory ensemble? The Jensen-Shannon divergence $J$~\cite{Cover:2006:Book} between these two trajectory ensembles  quantifies the average information gain, upon observing a trajectory randomly drawn with equal probability from either ensemble, about which ensemble it comes from~\cite{crooks11}. Essentially, this quantity tells us how weird a movie of a system's dynamics looks when played backwards.

Sufficiently slow processes remain near equilibrium, and provide a universal near-equilibrium result for irreversibility for a given amount of energy dissipation~\cite{feng08}. For a simple model of energy storage, irreversibility can be significantly increased above the near-equilibrium result by coupling storage of moderate-sized energy packets with moderate mechanical motion~\cite{brown16}, suggesting the energy and length scales of molecular machines may be constrained by a requirement for forward progress~\cite{brown17}.

\section*{\uppercase{{\bf Future prospects}}}

\subsection*{Stochastic driving}
\noindent Much of the theory above focuses on deterministic driving of a system, where a control parameter follows a fixed temporal schedule. Biomolecular machines do not typically experience an experimentalist deterministically changing a control parameter -- instead they operate autonomously, responding to the stochastic fluctuations of coupled nonequilibrium systems. Recent research~\cite{machta15,perunov16} has opened new vistas on the physical principles governing autonomous molecular machines driven by stochastic protocols.

\subsection*{Fundamental understanding of constraints for biology}
\noindent Although molecular machines are built and operate inside living cells, they are subject to physical constraints. In this article, we have outlined how molecular machines operate out of equilibrium to overcome viscous forces in the face of fluctuations. We also describe recent developments for efficient system control
and generation of irreversible dynamics. This emerging understanding of the statistical physics of driven microscopic processes points toward general principles for molecular machine operation.

\section*{\uppercase{{\bf Acknowledgments}}}
\noindent The authors thank their SFU Physics colleagues Nancy Forde, Chapin Korosec, Nathan Babcock, Steven Large, Aliakbar Mehdizadeh, and Joseph Lucero for feedback on the manuscript. 
This work was supported by a Natural Sciences and Engineering Research Council of Canada (NSERC) Discovery Grant (DAS), by funds provided by the Faculty of Science, Simon Fraser University through the President's Research Start-up Grant (DAS), and by support provided by WestGrid (www.westgrid.ca) and Compute Canada Calcul Canada (www.computecanada.ca).

\bibliographystyle{unsrt}
\bibliography{review}

\end{document}